\documentclass[superscriptaddress,twocolumn,aps,prb,showpacs]{revtex4-1}
\usepackage{graphicx}
\usepackage{amsbsy,amssymb,amsmath,bm,ulem}

\usepackage{color}
\usepackage{color}

\newcommand{\removed}[1]{}

\begin{document}

\title{Cascade dynamics of thermomagnetic avalanches in superconducting films with holes }

\author{J. I. Vestg{\aa}rden}
\affiliation{Department of Physics, University of Oslo, P. O. Box
1048 Blindern, 0316 Oslo, Norway}
\affiliation{Norwegian Defence Research Establishment (FFI), Kjeller, Norway}

\author{F. Colauto}
\affiliation{Departamento de F\'{i}sica, Universidade Federal de S\~{a}o Carlos, 13565-905, S\~{a}o Carlos, SP, Brazil}

\author{A.M.H. de Andrade}
\affiliation{Instituto de F\'{i}sica, Universidade Federal do Rio Grande do Sul, 91501-970, Porto Alegre, RS, Brazil}

\author{A.A.M. Oliveira}
\affiliation{Departamento de F\'{i}sica, Universidade Federal de S\~{a}o Carlos, 13565-905, S\~{a}o Carlos, SP, Brazil}
\affiliation{Instituto Federal de Educa\c{c}\~{a}o, Ci\^{e}ncia e Tecnologia de S\~{a}o Paulo, Campus Mat\~{a}o, SP, Brazil}

\author{W. A. Ortiz}
\affiliation{Departamento de F\'{i}sica, Universidade Federal de S\~{a}o Carlos, 13565-905, S\~{a}o Carlos, SP, Brazil}

\author{T. H. Johansen}
\affiliation{Department of Physics, University of Oslo, P. O. Box
1048 Blindern, 0316 Oslo, Norway}
\affiliation{Institute for Superconducting and Electronic
Materials, University of Wollongong, Northfields Avenue,
Wollongong, NSW 2522, Australia}





\begin{abstract}
The sub-microsecond dynamics of thermomagnetic avalanches in
superconducting films with non-conducting holes (antidots) is
considered. When such an avalanche reaches a hole, it is quickly
filled with magnetic flux, and often its rim becomes unstable and a
second avalanche is nucleated. In this work the time- and
space-resolved behavior of such cascading avalanche behavior is
determined using numerical simulations. Results are presented for
films with holes of different shape. It is found that holes with sharp
corners are those that most frequently create secondary avalanches,
and they tend to nucleate in corners. Magneto-optical imaging of Nb
films patterned with the same set of holes strongly supports the
numerical results.
\end{abstract}

\maketitle

In the critical-state of type-II superconductors,\cite{bean64} the
metastable system is susceptible to abrupt redistributions of magnetic
flux and electrical currents. For thin samples experiencing a
perpendicular magnetic field, even tiny field changes may cause large
amounts of flux to jump.  Typically, such an avalanche lasts less than
a microsecond,\cite{wertheimer67, leiderer93} and propagates in a
fingering structure, often strongly branched, as observed in films of
many materials\cite{ desorbo62, dolan73, duran95, johansen02,
  rudnev03, altshuler04-2, wimbush04, rudnev05, motta13,
  baziljevich14} using magneto-optical imaging (MOI). These dramatic
events are caused by an instability \cite{mints81} creating a runaway
of flux motion accompanied by local heating. Linear stability analysis
of the thermomagnetic model has explained why there exist onset
threshold values both in applied magnetic field and
temperature.\cite{mints96,denisov06,albrecht07} Numerical
simulations\cite{aranson01, aranson05,vestgarden11,vestgarden12-sr}
have confirmed that the equations representing the thermomagnetic
scenario of thin superconductors in perpendicular field indeed give
rise to the type of dendritic patterns found experimentally.

Interestingly, MOI observations have also revealed that the avalanches
show systematic new behavior when meeting macroscopic nonuniformities
in the sample. E.g., when a layer of normal metal covers part of the
superconducting film, an avalanche tends to change propagation
direction at the boundary between bare and coated
film.\cite{albrecht05,mikheenko13} A recent study of NbN films coated
with a Cu layer showed that the deflection of the avalanche branches
tend to follow Snell's law with a refraction index of 1.4, a number
also found to be the ratio of avalanche propagation velocities in the
bare and coated film.\cite{mikheenko15} Other experiments have shown
that inhomogeneities in the form of patterned holes in the
superconducting film also modify the avalanche behavior.  E.g., in
films with regular arrays of small holes (antidots), one finds strong
guidance of the avalanche
branches.\cite{vlasko-vlasov00,menghini05,motta11,motta14} How
avalanches are affected by large holes has also been investigated.  In
the work Ref.~\onlinecite{olsen07} the process of flux jumping into
the central hole of a circular planar ring of MgB$_2$ was studied,
revealing major redistribution of the shielding
currents. Recently,\cite{colauto13} macro-holes in Nb films were
considered as potential traps, or stop holes, serving to limit the
spatial extension of thermomagnetic avalanches.

The present work aims to reveal the utra-fast cascade dynamics of
avalanches propagating in superconducting films patterned with
macro-holes. Space- and time-resolved results from quantitative
numerical simulations for samples with holes of various shapes are
presented, and compared with avalanche behavior observed in a
similarly patterned superconducting film.

\begin{figure}[b]
  \centering
  \includegraphics[width=8cm]{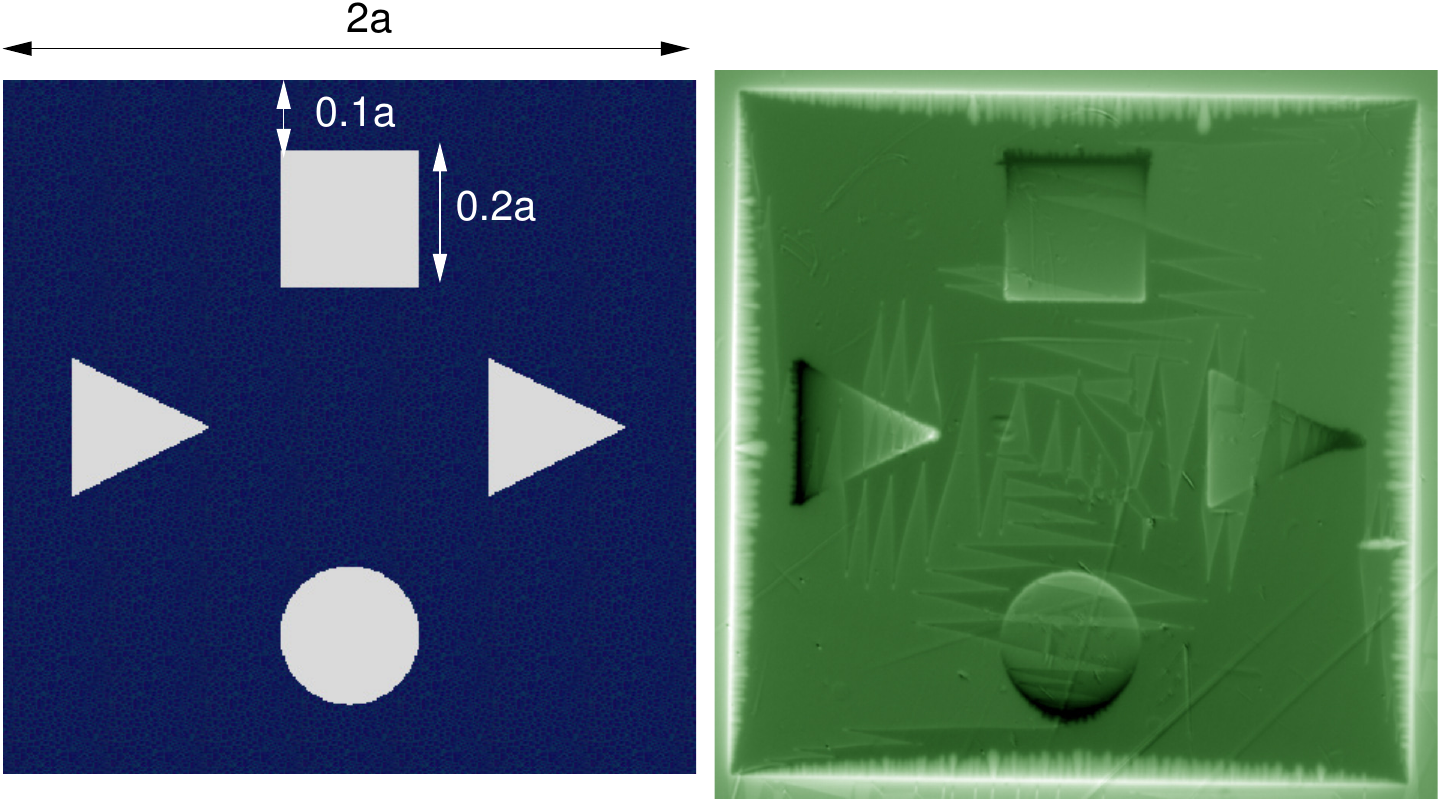} 
  \caption{
    \label{fig:sample}
    (left) Sketch of the sample, a square film with four holes. (right)
    Magneto-optical image of the perpendicular flux density, $B_{\rm z}$, 
in a Nb film at an applied field below the
   onset threshold for avalanche activity.  
  }
\end{figure}

MOI experiments were conducted on a square Nb film of sides
$2a=2.5$~mm and thickness $d=200$~nm.  Using optical lithography the
film was patterend with holes shaped as a square, a circle, and two
triangles, see Fig.~\ref{fig:sample} (left). Each hole has a dimension
$0.2a$, and is placed adistance $0.1a$ from the nearest edge.  To
visualize the flux density a ferrite-garnet film with large Faraday
coefficient\cite{helseth02-3} was placed directly on the sample, and
after cooling down in an optical cryostat, polarized light microscopy
was performed.

The Nb film was first zero-field-cooled to $5~$K, a temperature
sufficiently low for avalanches to occur. When a transverse magnetic
field is gradually applied, flux begins to penetrate, and when
reaching 2.5~mT one observes the magneto-optical image shown in
Fig.~\ref{fig:sample} (right). The image was recorded with slightly
uncrossed polarizers in the MOI setup, allowing to discriminate
between flux into or out of the image plane. The central part of the
film, which is in the Meissner state serves as reference color
representing $B_z=0$. The bright rim around the sample perimeter shows
the applied field piling up outside the diamagnetic film. The image
also shows a shallow flux penetration from the edge, in a pattern
typical for critical-state behavior. Note that the holes are clearly
visible in the image. All four holes display a dark side towards the
sample edge, and a bright side towards the center. This bipolar
(flux-antiflux) feature is caused by the additional field created by
the shielding currents as they are forced to circumvent the
holes.\cite{baziljevich96-2}

\begin{figure}[t]
  \centering
  \includegraphics[width=4.0cm]{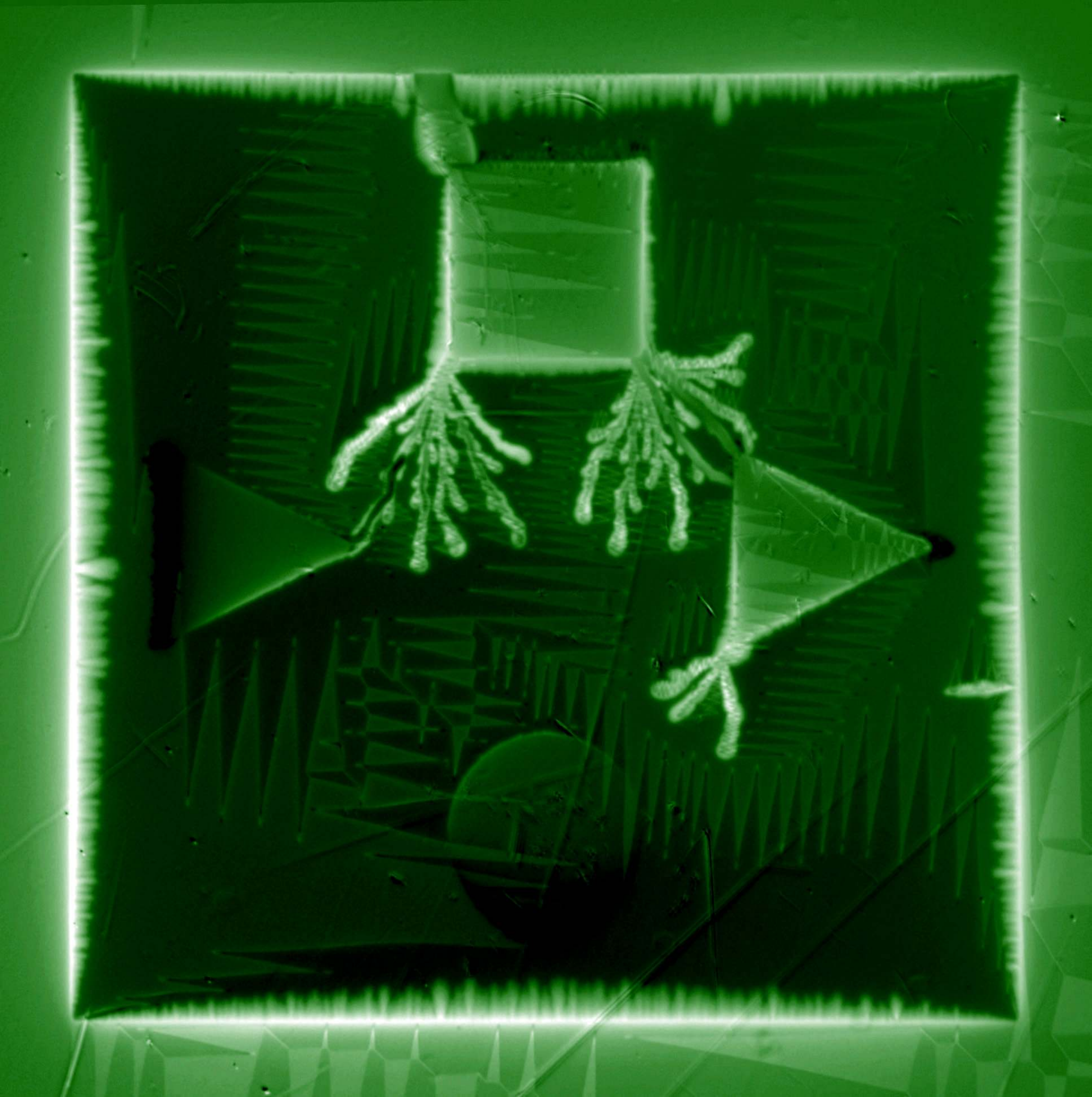}
  \includegraphics[width=4.0cm]{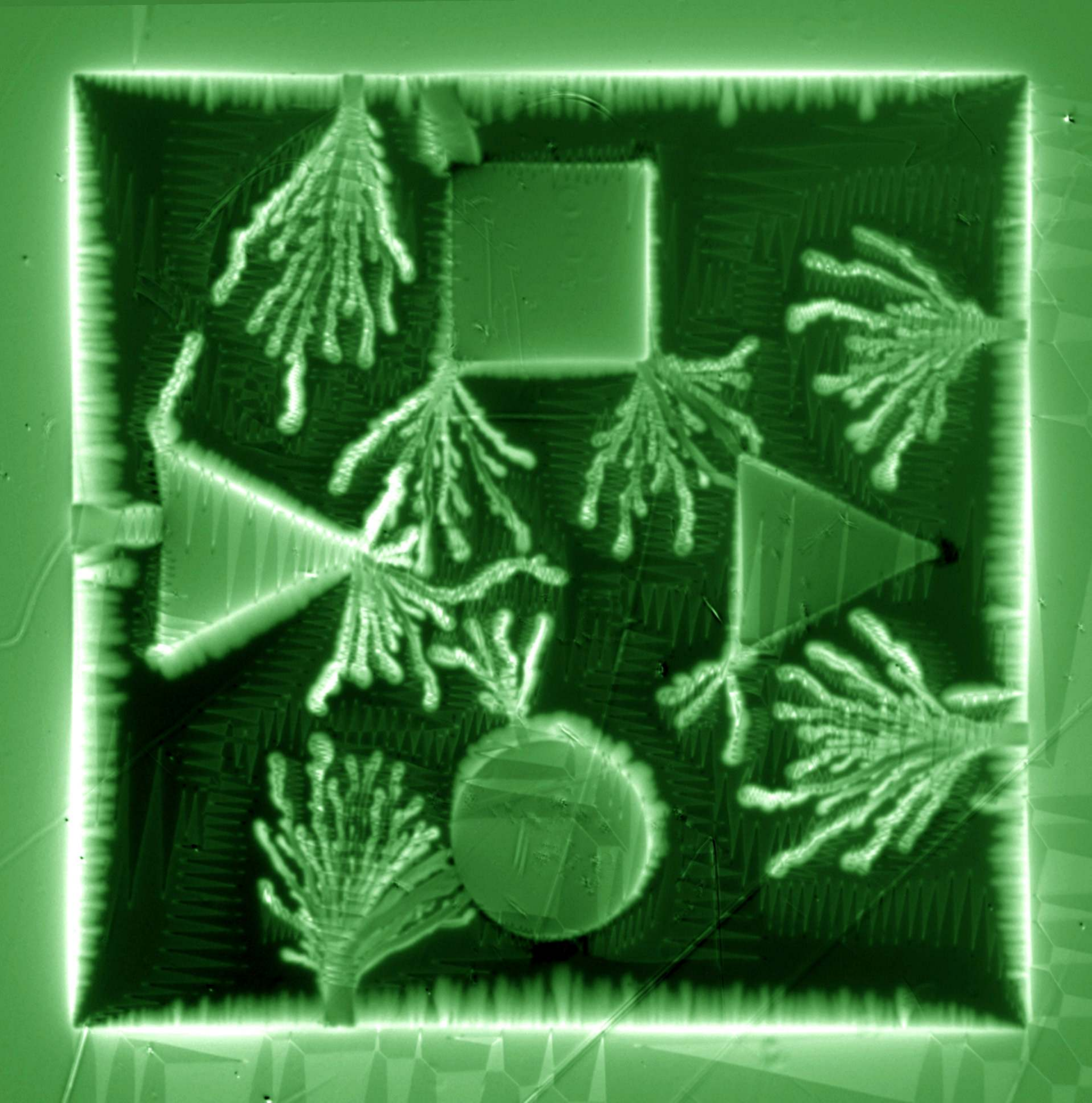} 
  \caption{
    \label{fig:MO}
 Magneto-optical images of  $B_{\rm z}$ in a Nb film cooled to 5~K, and then exposed to perpendicular applied fields of $2.8~$mT (left) and $4.35~$mT (right). The branching patterns in the flux distribution are  traces of the flux avalanches.
  }
\end{figure}

When the applied field reaches 2.8~mT the first avalanche event
occurs, resulting in the final flux distribution shown in
Fig.~\ref{fig:MO} (left). The avalanche started from the upper edge,
and from there invaded the square hole through the thick finger seen
to bridge the two regions. Then, from the two inner corners of the
square hole secondary events started, resulting in two highly branched
avalanches.  Both of them reached their nearest triangular hole, which
by their bright constrast, are seen to have received considerable
amounts of flux. Then, as a final step in this cascade of events, the
triangular hole on the right side releases its magnetic pressure in an
avalanche consisting of 3 branches.

By increasing the field further, even more avalanches appear, and
Fig.~\ref{fig:MO} (right) shows the flux distribution at the field of
4.35~mT. Here, all the holes have received flux, and one sees that
some avalanches, e.g., both events starting from the right sample edge
do not feed flux into any hole. Continued field ramping adds more and
more avalanches, and the complexity in the overall pattern grows.

\begin{figure}[t]
  \centering
  \includegraphics[width=8cm]{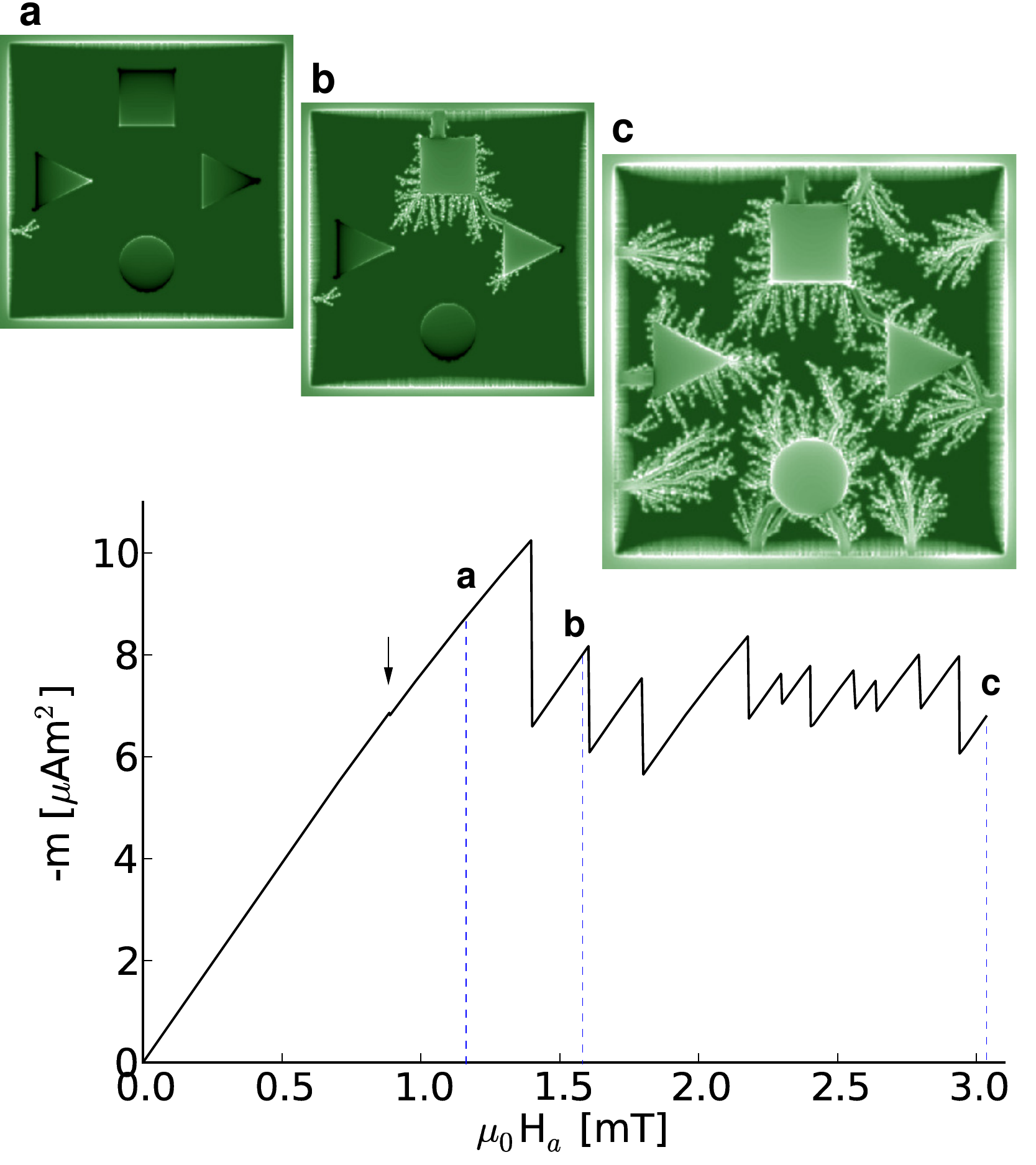} 
  \caption{\label{fig:moment}
    Magnetic moment as  function of applied field 
    obtained by numerical simulation. Each drop in the curve corresponds 
    to an avalanche event. The arrow markes the first avalanche. 
The images (a)-(c) show the $B_{\rm z}$-distribution
    at applied fields of  1.2,  1.55 and  3.05 mT, respectively. 
  }
\end{figure}

\begin{figure*}[t]
  \centering
  \includegraphics[width=14.7cm]{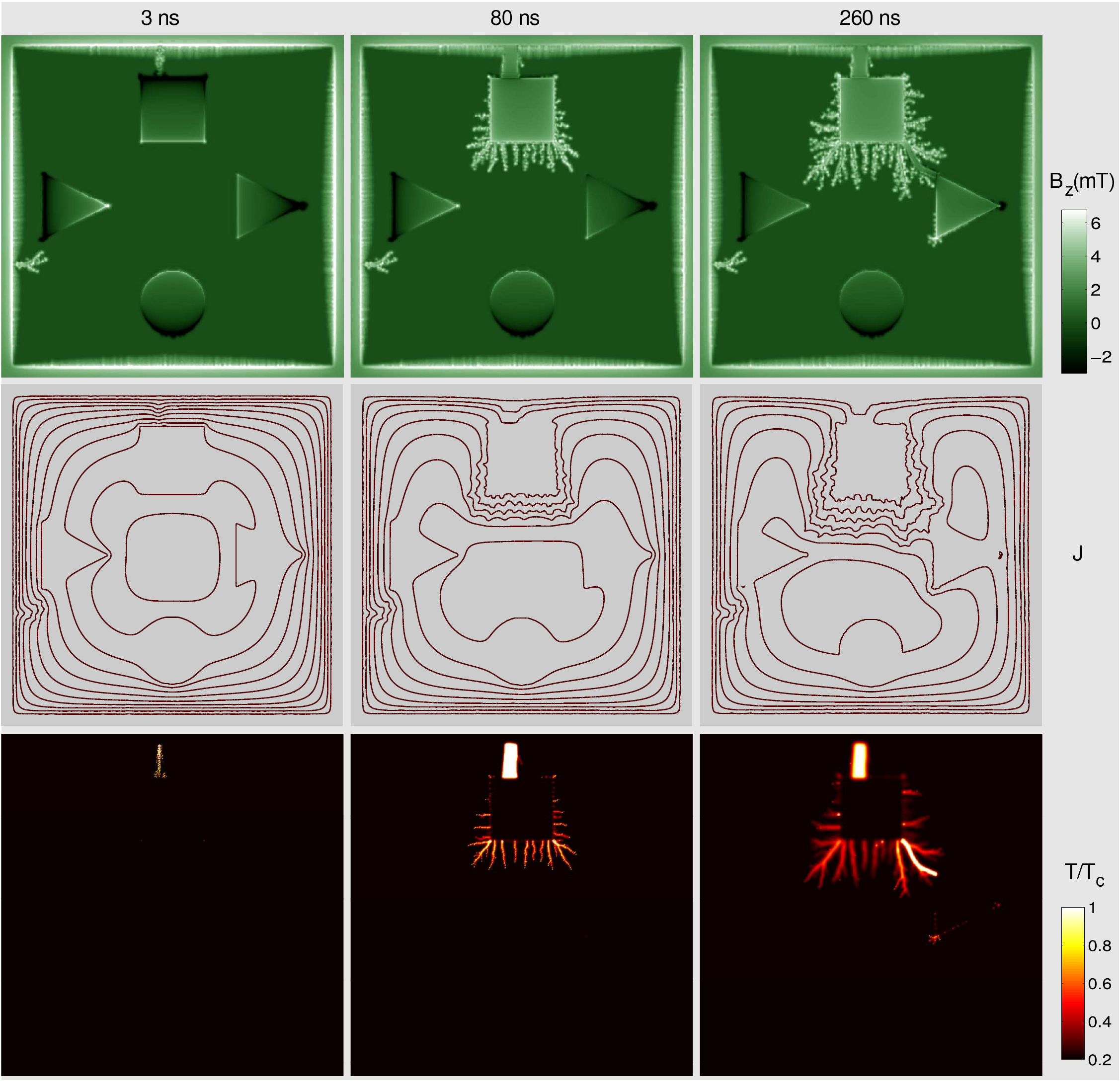} 
  \caption{\label{fig:HJT}
    Numerical simulations of the avalanche  occuring at $\mu_0H_{\rm a} = 1.4$~mT. 
    Horizontal rows  show the flux distribution,  $B_{\rm z}$ (upper),  
    current stream lines (middle) and temperature (lower) at times
    3, 86 and 260~ns after nucleation of the event.
  }
\end{figure*}

In the following we present results from numerical simulations of such
sequences of events, thus allowing deeper insight into this ultrafast
cascading behaviour. 

The electrodynamical part of the problem is solved by time-integration
of the Maxwell equations
\begin{equation}
  \dot {\mathbf B} = -\nabla\times \mathbf E,~~
  \nabla\times \mathbf H = \mathbf J\delta(z),~~
  \nabla\cdot \mathbf B = 0,
\end{equation}
with $\mu_0\mathbf H = \mathbf B$ and $\nabla\cdot \mathbf J = 0$,
using implicit boundary conditions for edges and holes,
as described in Ref.~\onlinecite{vestgarden13-fftsim}.
The material characteristics of the sample is 
the conventional power-law $E-J$ relation
\begin{equation}
  \mathbf E =  \rho_0\mathbf J/d  \left\{
  \begin{array}{ll}
    1, & |\mathbf J| > dj_{\rm c} \mbox{ or } T > T_{\rm c}, \\
    \left(|\mathbf J|/dj_{\rm c}\right)^{n-1}, & \mbox{otherwise}.
  \end{array}
  \right.
  \label{rho}
  , 
\end{equation}
where $T$ is the local temperature, $j_{\rm c}$ is the 
critical current, and $n$ is
the creep exponent. The temperature dependencies are 
$\label{jcT} j_{\rm c}=j_{\rm c0}(1-T/T_{\rm c})$ and $n = n_1T_{\rm c}/T$.
The local heat is found by time-integration of 
\begin{equation}
  \label{dotT}
  c\dot T = \kappa \nabla^2 T - h \left(T-T_0\right) /d + JE/d
  ,
\end{equation}
where $c$ is the specific heat, $\kappa$ is the heat conductivity,
$h$ is the coefficient of heat transfer to the substrate,
and $T_0$ is the substrate temperature..

The simulations use the following parameters:
$T_{\rm c}=9.2~$K,  $\rho_0$ = 5~$\cdot 10^{-9}~\Omega$m,
$j_{\rm c0} = 1.2 \cdot 10^{11}$~A/m$^2$,
$\kappa=\left[20\right.$~W/Km$\left.\right](T/T_{\rm c})^3$,
$c = \left[3\cdot 10^4\right.$ ~J/Km$^3$$\left.\right](T/T_{\rm c}) ^3$,
and
$h= \left[10^4\right.$~W/Km$^2$$\left.\right](T/T_{\rm c}) ^3$.
We set $n_1=20$ and restrict $n(T)$ to $ n\leq 100$.
The sample dimensions are $a=1.3~$mm and $d=100~$nm, and
the applied magnetic field is ramped at the constant rate of
$\dot H_{\rm a}= 2.4\cdot 10^{-6}j_{\rm c0}\rho_0/a\mu_0$,
from initially $\mathbf B_z = 0$ and $T=T_0\equiv 0.2 T_{\rm c}$.

As the applied field gradually increases, shielding currents are
induced in the sample, giving it a magnetic moment,
$m$. Figure~\ref{fig:moment} displays the moment as function of field,
and shows that at low fields the magnitude of $m$ increases
monotonously, as expected from the growing induced currents.  Soon,
the sheet current near the edge reaches the critical value 
$J_{\rm c}$, and magnetic flux starts to enter the sample along the rim.  At
the field of 0.8~mT a minor avalanche strikes near one of the
triangular holes. The event is clearly visible in the flux
distribution of image (a), but is hardly noticeable in the magnetic
moment curve.

The next event occurs at $\mu_0H_{\rm a}=1.4$~mT, and is a much bigger
avalanche, as $m$ here makes a large drop. Image (b) shows that this
avalanche alters the flux distribution significantly, and hasfilled
both the square and triangular hole on the right side with sizable
amounts of flux. Thereafter, the magnetization curve shows that
avalanche events continue to appear with irregular intervals and
magnitudes.  The simulation was ended at the applied field of
$3.05~$mT, with the flux distributed as seen in image (c). During this
field ramp a total of 11 avalanches took place, and the sequence of
images of $B_z$ reproduces most features of the shown experimental
images.

To illustrate the time evolution of an avalanche, the
Fig.~\ref{fig:HJT} shows snapshots from the simulations, recorded at
three moments in time after nucleation of the event at 
$\mu_0H_{\rm a}=1.4$~mT. The figure presents maps of the flux density 
$B_{\rm z}$, the stream lines of the sheet current $J$, and the local
temperature $T$. From the left column, one sees that already after 3
ns, a narrow channel of moving flux, accompanied by elevated
temperature, is connecting the square hole to the flux reservoir
outside the edge.

\begin{figure}[t]
  \centering
  \includegraphics[width=7.5cm]{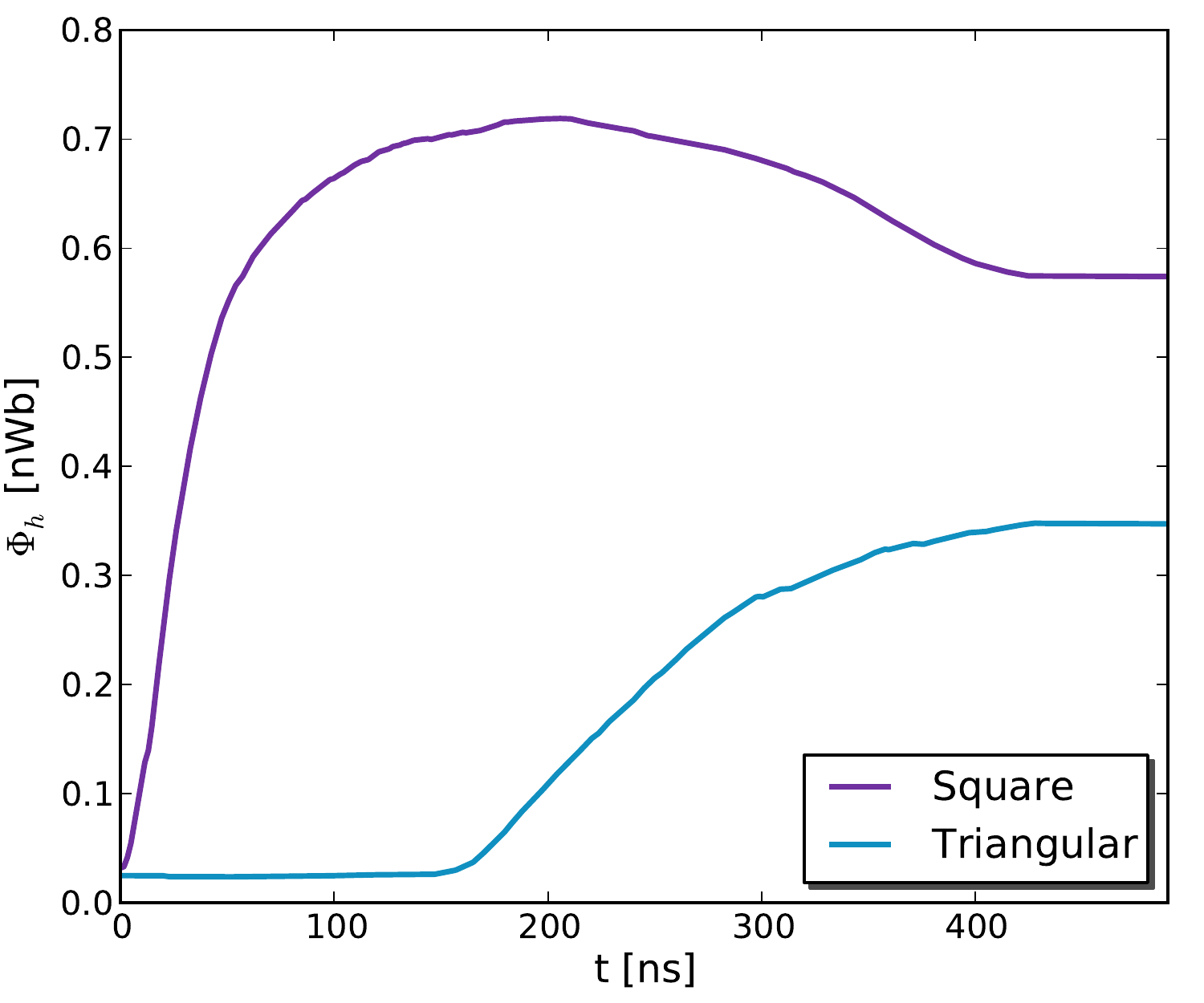} \\
  \caption{\label{fig:phi-d0000}
    Total flux in the square and the triangular hole 
    as functions of time during the avalanche at $\mu_0 H_a = 1.4~$mT.  
  }
\end{figure}

After 80 ns (middle column) this channel has grown in width, and the
flux traffic has increased significantly. The $T$-map shows that in
the entire channel the temperature now rises well above $T_c$. Via
this opening the square hole becomes filled with flux, which in turn
nucleates secondary avalanches from the edge of the hole. The
corresponding $J$-map confirms that a massive rearrangement of
currents is taking place. Actually, at this stage the hole is more
part of the outside than the inside of the superconductor, as far as
shielding is concerned.

After 260 ns (right column) the secondary avalanches have penetrated
deeper. In fact, one branch has reached another hole, namely the
triangular one on the right side. That hole is now in being filled
with flux via the narrow channel appearing as very bright in the
$T$-map.

A more quantitative illustration of the interaction between avalanches
and holes is presented in Fig.~\ref{fig:phi-d0000}.  The two graphs
display the total flux $\Phi_{\rm h}$ entering the square and
triangular hole as functions of time during the avalanche occurring at
$\mu_0 H_a = 1.4~$mT. First, flux is rapidly injected into the square
hole with almost constant rate, while the flux inside the triangular
hole stays constant. At around $t = 200$~ns the flux in the square
hole reaches a maximum. This occurs nearly at the same time as the
flux in the triangular hole begins to increase.  After 400~ns both
curves reach a plateau as the avalanche comes to rest.

In the initial stage of the avalanche, the rate of change of the flux
in the square hole is $\dot \Phi_{\rm h} = 0.012$~Wb/s.  This
represents the flux entering the hole while the channel to the
external rim is heated above $T_c$. From Fig.~\ref{fig:HJT} one can
estimate that the channel width is $l=25~\mu$m, which means that in
the channel the electric field amounts to $E=\dot \Phi_{\rm h}
/l=500$~V/m.  For comparison, an estimate of the electric field at the
sample's outer edge created by the regular flux penetration gives $E=
\mu_0\dot H_aa=1.4~$mV/m.  Note that in the present model no heat
flows in the substrate below the holes. This suggests that the
electrodynamics alone can lead to avalanche cascades between holes in
a superconducting film.

In summary, the present work has shown that thermomagnetic avalanches
in superconducting films tend to be attracted by holes in the
film. The numerical simulations reveal the detailed dynamics at the
various stages of an avalanche as it propagates from one hole to
another. Typically, a hole becomes filled with flux via a narrow
normal-state channel created between the film's external edge and the
hole. It is found that this often lead to secondary avalanches, which
spread to neighboring holes or parts of the film not yet penetrated by
flux. Holes with sharp corners, like square and triangular holes, are
those that most frequently create avalanche cascades.

The samples were grown in Laborat\'{o}rio de Conforma\c{c}\~{a}o
Nanom\'{e}trica (LCN-IF-UFRGS), and the lithography was made in
Laborat\'{o}rio de Microfabrica\c{c}\~{a}o (LMF/LNNano/CNPEM).
The work was financially supported by the Brazilian funding agencies FAPESP
and CNPq, and the program Science without Borders, as well as the 
CAPES-SIU-2013/10046 project "Complex fluids in confined environments".

%

\end{document}